\documentclass[10pt]{article}
\pdfoutput=1
\usepackage[T1]{fontenc}
\usepackage{titling}
\usepackage{hyperref}

\usepackage{tocloft}  
\usepackage[utf8]{inputenc}
\usepackage{amsmath}
\numberwithin{equation}{section} 
\usepackage{amsfonts}
\usepackage{amssymb}
\usepackage[version=4]{mhchem}
\usepackage{stmaryrd}

\usepackage[export]{adjustbox}
\usepackage{graphicx}
\graphicspath{ {./images/} }
\usepackage{geometry}
\usepackage{fancyhdr}  

\usepackage{caption}
\usepackage{cite}

\usepackage[T1]{fontenc}

\usepackage{listings}
\usepackage{xcolor}

\usepackage{titling}
\pretitle{\begin{center}\LARGE}
\posttitle{\end{center}\thispagestyle{empty}} 

\title{Grid Modeling of Renewable Energy}
\author{Dr. Sohail Khan}

\date{\today}

\lstdefinestyle{matlabstyle}{
	backgroundcolor=\color{lightgray},    
	basicstyle=\ttfamily\small,           
	keywordstyle=\color{blue},            
	commentstyle=\color{yellow},           
	stringstyle=\color{orange},           
	numberstyle=\tiny\color{gray},        
	rulecolor=\color{black},
	breaklines=true,                      
	frame=single,                         
	language=Matlab                       
}

\begin{document}

\maketitle
\tableofcontents
\newpage

\section*{License}
\noindent
This document is licensed under the \textbf{Creative Commons Attribution-NoDerivs 4.0 International License}. 
To view a copy of this license, visit \href{https://creativecommons.org/licenses/by-nd/4.0/}{https://creativecommons.org/licenses/by-nd/4.0/}. \\
Under this license, you are free to share, copy, and distribute the material in any medium or format for educational purpose only,  as long as you give appropriate credit by citing the lecture notes from ArxiV. 
\textbf{No derivatives or adaptations of this material are allowed.}

\section*{Preamble}
The lecture notes focus on the modeling of power system components with focus on modeling of components representing renewable energy generation, energy storage and related enabling technologies.
The simulation testing shall be performed using Matpower \cite{zimmerman2010matpower} and Pandapower \cite{thurner2018pandapower}.  

\section{Power System Network Modeling}
This section introduces the core elements of network modeling in power systems.

\subsection{Branches}
The transmission \& distribution lines (does not include DC transmission or distribution lines),  transformers, and phase shifters are typically modeled with a common branch model. 
The common branch model is based on the standard $\pi$ transmission line model describe in \cite{saadat1999power} in series with an ideal phase-shifting transformer. 

\begin{figure} [htb]
	\centering
	\includegraphics[trim=0cm 0cm 0cm 0cm, width=0.8\textwidth]{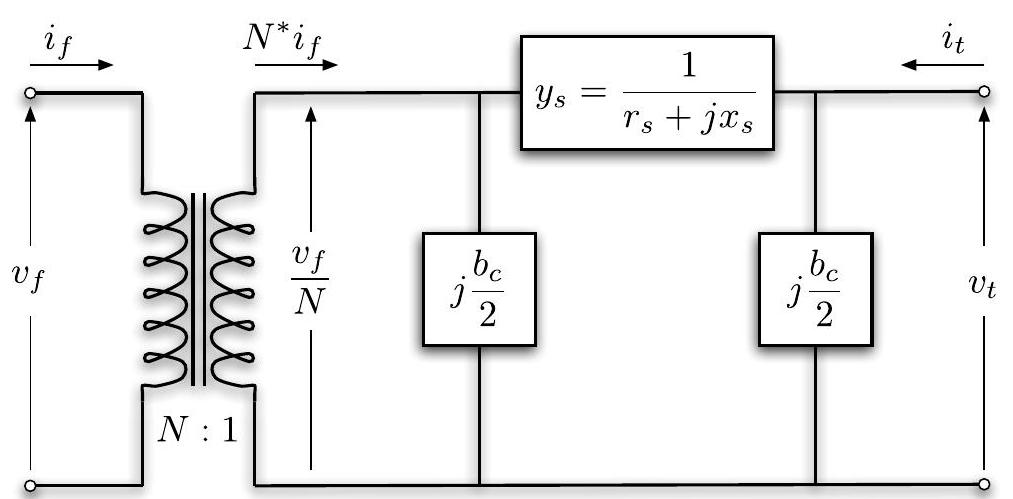}
	\caption{Branch model based on $\pi$ model of transmission lines.}
	\label{figure: branch_model}
\end{figure}

A typical transmission/distribution line has a series impedance $z_{s}=r_{s}+j x_{s}$ representing its resistance and reactance (admittance $y_s$ is reciprocal of impedance $z_s$)  \& a total charging susceptance $b_{c}$ split in two as from and to ends of the line, as shown in Fig. \ref{figure: branch_model}. The ideal phase shifting transformer is modeled as tap ratio of magnitude $\tau$ and phase shift angle $\theta_{\text {shift}}$, is located at the from end of the branch. 

Following is the brief summary of the $\pi$ model and derivation of the branch admittance matrix representing it:	
\begin{itemize}
	\item A series impedance $z_{s} = r_{s} + jx_{s}$ and $y_s=\frac{1}{z_{s}}$, represent the line impedance and admittance, respectively. 
	\item Two shunt admittance $Y_{\text{shunt}} = j \frac{b_{c}}{2}$ at the sending and receiving ends of the line, representing the capacitive charging current of the line.
	\item A phase-shifting transformer introduces a phase shift $\theta_{\rm shift}$ between the sending and receiving ends of a transmission line. It is typically modeled as a complex turn ratio $\tau$, where:
	\begin{equation}	
		\tau = |\tau| e^{j \theta_{\rm shift}}
	\end{equation}
	The transformer affects both the voltage magnitude and phase angle across the line. The ideal line will have zero angular shift and $|\tau|=1$. The branch admittance matrix must account for the impact of this phase shift.
\end{itemize}
	
\newpage
\textbf{Derivation of the Branch Admittance Matrix}

Let the sending end and receiving end voltages be $v_f$ and $v_t$, respectively.
The respective complex current injections $i_{f}$ and $i_{t}$ at the from and the to ends of a branch can be expressed in terms of the $2 \times 2$ branch admittance matrix $Y_{b r}$ and the corresponding terminal voltages $v_{f}$ and $v_{t}$, based on Ohms law :
\begin{equation}
	\left[\begin{array}{l}
		i_{f} \\
		i_{t}
	\end{array}
	\right]=Y_{br}
	\left[\begin{array}{l}
		v_{f} \\
		v_{t}
	\end{array}\right]
	=
	\begin{pmatrix}
		Y_{ff} & Y_{ft} \\
		Y_{tf} & Y_{tt}
	\end{pmatrix}
	\left[\begin{array}{l}
		v_{f} \\
		v_{t}
	\end{array}\right]
	\label{eq:from_to_branch_equation}
\end{equation}

\begin{equation}
\begin{pmatrix}
	i_f \\
	i_t
\end{pmatrix}
= 
\begin{pmatrix}
	Y_{ff} & Y_{ft} \\
	Y_{tf} & Y_{tt}
\end{pmatrix}
\begin{pmatrix}
	v_f \\
	v_t
\end{pmatrix}
\end{equation}
where $Y_{ff}$, $Y_{ft}$, $Y_{tf}$, and $Y_{tt}$ are the elements of the branch admittance matrix.

\textbf{Step 1: Admittance Without Phase Shift}

First, we consider the admittance of the transmission line without the phase shift. The admittance matrix for the $\pi$ model is:
\begin{equation}
Y_{br}  = \begin{pmatrix}
	y_{s} + j \frac{b_c}{2} & -y_{s} \\
	-y_{s} & y_{s} + j \frac{b_c}{2}
\end{pmatrix}
\end{equation}
For derivation, refer to of Section 6.2 of \cite{saadat1999power} stating that the diagonal element is the sum of admittance connected to a node also referred as self-admittance and off-diagonal element is equal to the negative of admittance between nodes. 

\textbf{Step 2: Effect of the Phase Shift}

With the phase-shifting transformer, the current-voltage relationship is modified by the complex turn ratio $\tau$. 
The sending-end voltage $v_f$ impacted by the $tau$ is delivered to the receiving-end voltage $v_t$:
\begin{equation}
 v_t = v_f \times \tau
\end{equation}
Substituting this into the admittance matrix, the modified admittance matrix becomes (refer to Section 6.7 of \cite{saadat1999power} for derivation):
\begin{equation}
Y_{br} = \begin{pmatrix}
	\frac{y_{s} + j \frac{b_c}{2}}{|\tau|^2} & -\frac{y_{s}}{\tau^*} \\
	-\frac{y_{s}}{\tau} & y_{s} + j \frac{b_c}{2}
\end{pmatrix}
\end{equation}
where $\tau^*$ is the complex conjugate of $\tau$.
This matrix can now be used in power flow calculations and other power system studies where the effect of the phase shift must be considered.

Denoting the series admittance element in the $\pi$ model by $y_{s}=1 / z_{s}$, the branch admittance matrix can be written as

\begin{equation}
	Y_{b r}=\left[\begin{array}{cc}
		\left(y_{s}+j \frac{b_{c}}{2}\right) \frac{1}{\tau^{2}} & -y_{s} \frac{1}{\tau e^{-j \theta_{\text {shift }}}}  \\
		-y_{s} \frac{1}{\tau e^{j \theta_{\text {shift }}}} & y_{s}+j \frac{b_{c}}{2}
	\end{array}\right]
	\label{equation:branch admittance matrix}
\end{equation}

\subsection{Generator}
A generator is modeled as a complex power injection at a specific bus. The injection by generator $i$ is written as

\begin{equation}
s_{g}^{i}=p_{g}^{i}+j q_{g}^{i} 
\end{equation}

Let $S_{g}=P_{g}+j Q_{g}$ be the $n_{g} \times 1$ vector of these generator injections. 
A sparse $n_{b} \times n_{g}$ generator connection matrix $C_{g}$ can be defined such that its $(i, j)^{\text {th }}$ element is 1 if generator $j$ is located at bus $i$ and 0 otherwise. The $n_{b} \times 1$ vector of all bus injections contributed by the generator is then expressed as

\begin{equation}
S_{g, \text {bus}}=C_{g} \cdot S_{g} \text {. } 
\end{equation}

A generator with a negative injection can also be used to model a dispatch-able load.

\subsection{Loads}
A constant-power load is modeled as a specified active (real) and reactive power consumed at a bus. The load at bus $i$ is written as

\begin{equation}
s_{d}^{i}=p_{d}^{i}+j q_{d}^{i} 
\end{equation}

Let $S_{d}=P_{d}+j Q_{d}$ be the $n_{b} \times 1$ vector of complex load consumption at the buses. 
The entries in this vector can also take on negative quantities to represent fixed (e.g., distributed) generation. Note that, dispatchable loads are modeled as negative generators and appear as negative values in $S_{g}$. 

\subsection{Shunt Elements}
A shunt-connected element, such as a capacitor or an inductor, is modeled as a fixed impedance to ground at a bus. The admittance of the shunt element at bus $i$ is written as

\begin{equation}
y_{s h}^{i}=g_{s h}^{i}+j b_{s h}^{i} 
\end{equation}

Let $Y_{s h}=G_{s h}+j B_{s h}$ denote the $n_{b} \times 1$ vector of shunt admittance at all buses. 
A shunt element can also be used to represent a constant impedance load.

\subsection{Network Equations}
For a network with $n_{b}$ buses, all constant-impedance elements of the system model are incorporated into the complex $n_{b} \times n_{b}$ bus admittance matrix $Y_{\text {bus }}$. The complex nodal current injections $I_{\text {bus }}$ are linearly related to the complex node voltages $V$ through the bus admittance matrix:

\begin{equation}
I_{\text {bus }}=Y_{\text {bus }} V 
\label{eq:kirchoff_current_low}
\end{equation}

\subsubsection{Relationship between Bus Admittance Matrix and Branch Admittance Matrix}
The bus admittance matrix $Y_{bus}$, plays a central role in power flow analysis. It provides a compact representation of how the buses in a power system are interconnected through transmission lines and other electrical components. The branch admittance matrix, denoted as $Y_{b r}$, represents the admittances of individual transmission lines (branches) connecting two buses.

The relationship between the bus admittance matrix and the branch admittance matrix can be expressed using network topology and Kirchhoff’s laws. The bus admittance matrix is constructed from the branch admittance matrix by summing the admittances of all branches connected to a bus. The diagonal elements of the bus admittance matrix represent the sum of the admittances of the branches connected to that bus, while the off-diagonal elements represent the negative of the branch admittance between two buses.

For a system with $n$ buses and $m$ branches, the bus admittance matrix $Y_{bus}$ is derived from the branch admittance matrix $Y_{b r}$ using the incidence matrix $\mathbf{A}$ of the network. This can be expressed as:

\begin{equation}
	Y_{bus} = \mathbf{A}^T Y_{b r} \mathbf{A}
\end{equation}

where:
\begin{itemize}
	\item $\mathbf{A}$ is the incidence matrix that represents the connection between buses and branches.
	\item $Y_{b r}$ is the diagonal matrix of branch admittances.
\end{itemize}

Similarly, for a network with $n_{l}$ branches, the $n_{l} \times n_{b}$ system branch admittance matrices $Y_{f}$ and $Y_{t}$ respectively relate the bus voltages to the $n_{l} \times 1$ vector $I_{f}$ and $I_{t}$ of branch currents at the from and the to ends of the branches:
\begin{align}
I_{f} & =Y_{f} V  \\
I_{t} & =Y_{t} V 
\end{align}

The corresponding complex power injections can be computed as functions of the complex bus voltages $V$ :
\begin{align}
S_{\text {bus }}(V) & =[V] I_{\text {bus }}^{*}=[V] Y_{\text {bus }}^{*} V^{*} 
\end{align}

The nodal bus injections are then matched to the injections due to the generator and loads, forming the AC nodal power balance equations. Expressed as a function of the complex bus voltages and generator injections, they are written in complex matrix form as
\begin{equation}
g_{S}\left(V, S_{g}\right)=S_{\mathrm{bus}}(V)+S_{d}-C_{g} S_{g}=0 
\end{equation}

\subsection{DC Modeling}
The DC formulation is based on the same parameters, but with the following three additional simplifying assumptions.
\begin{itemize}
  \item The branches can be considered lossless. In particular, branch resistances $r_{s}$ and charging capacitance $b_{c}$ are negligible:
\end{itemize}
\begin{equation}
y_{s}=\frac{1}{r_{s}+j x_{s}} \approx \frac{1}{j x_{s}}, \quad \text { and } \quad b_{c} \approx 0 
\end{equation}
\begin{itemize}
  \item All bus voltage magnitudes are close to 1 p.u., i.e.,
\end{itemize}
\begin{equation}
v_{i} \approx e^{j \theta_{i}} 
\label{eq:second_assumption}
\end{equation}
\begin{itemize}
  \item Voltage-angle differences across branches are small enough that
\end{itemize}
\begin{equation}
\sin \left(\theta_{f}-\theta_{t}-\theta_{\text {shift }}\right) \approx \theta_{f}-\theta_{t}-\theta_{\text {shift }} 
\label{eq:dc_branch_equation}
\end{equation}

Invoking the assumptions regarding the branch parameters, the branch admittance matrix (Eq. \ref{equation:branch admittance matrix}) approximates to:

\begin{equation}
Y_{b r} \approx \frac{1}{j x_{s}}\left[\begin{array}{cc}
\frac{1}{\tau^{2}} & -\frac{1}{\tau e^{-j \theta_{\text {shift }}}}  \\
-\frac{1}{\tau e^{j \theta_{\text {shift }}}} & 1
\end{array}\right]
\end{equation}

\section{Power Flow}\label{section:power_flow}
The standard power flow or load-flow problem involves solving for the set of voltages and flows in a network corresponding to a specified pattern of load and generation. It involves solving a set of equations of the form constructed by expressing a subset of the nodal power balance equations as functions of unknown voltage quantities.
\begin{equation}
g(x)=0 
\end{equation}

\subsection{AC Power Flow}
By convention, a single generator bus or connection with the external power system network is typically chosen as a reference bus (or more commonly referred as slack bus). 
The slack bus serve the roles of both a voltage magnitude and angle reference. 
The real and reactive power generation or exchange at the slack bus is taken as unknown to avoid over specifying the problem. 

The remaining generator buses are typically classified as PV buses, with the values of voltage magnitude and generator real power injection given. 
Since the loads $P_{d}$ and $Q_{d}$ are also given, all non-generator buses are classified as PQ buses, with real and reactive injections fully specified. 
Let $\mathcal{I}_{\text {ref }}, \mathcal{I}_{\mathrm{PV}}$ and $\mathcal{I}_{\mathrm{PQ}}$ denote the sets of bus indices of the reference bus, PV buses and PQ buses, respectively. 

In the traditional formulation of the AC power flow problem, the power balance equation is split into its real and reactive components, expressed as functions of the voltage angles $\Theta$ and magnitudes $V_{m}$ and generator injections $P_{g}$ and $Q_{g}$, where the load injections are assumed constant and are given as:

\begin{align}
& g_{P}\left(\Theta, V_{m}, P_{g}\right)=P_{\mathrm{bus}}\left(\Theta, V_{m}\right)+P_{d}-C_{g} P_{g}=0
\label{eq:real_power_balance_AC}  \\
& g_{Q}\left(\Theta, V_{m}, Q_{g}\right)=Q_{\mathrm{bus}}\left(\Theta, V_{m}\right)+Q_{d}-C_{g} Q_{g}=0 
\label{eq: reactive_power_balance_AC}
\end{align}

For the AC power flow problem, the function $g(x)$ is formed by taking the left-hand side of the real power balance equations (Eq.~\ref{eq:real_power_balance_AC}) for all non-slack buses and the reactive power balance equations (Eq.~\ref{eq: reactive_power_balance_AC}) for all PQ buses and plugging in the reference angle, the loads and the known generator injections and voltage magnitudes:

\begin{equation}
	g(x)=\left[\begin{array}{cl}
		g_{P}^{i}\left(\Theta, V_{m}, P_{g}\right)  \\
		g_{Q}^{j}\left(\Theta, V_{m}, Q_{g}\right)
	\end{array}\right] \quad 
	\begin{aligned}
		& \forall i \in \mathcal{I}_{\mathrm{PV}} \cup \mathcal{I}_{\mathrm{PQ}} \\
		& \forall j \in \mathcal{I}_{\mathrm{PQ}}
	\end{aligned}
\end{equation}

The vector $x$ consists of the remaining unknown voltage quantities, namely the voltage angles at all non-reference buses and the voltage magnitudes at PQ buses:

\begin{equation}
	x=\left[\begin{array}{c}
		\theta^{i}  \\
		v_{m}^{j}
	\end{array}\right] \quad 
	\begin{aligned}
		& \forall i \notin \mathcal{I}_{\text {ref }} \\
		& \forall j \in \mathcal{I}_{\mathrm{PQ}}
	\end{aligned}
\end{equation}

This yields a system of nonlinear equations with $n_{p v}+2 n_{p q}$ equations and unknowns, where $n_{p v}$ and $n_{p q}$ are the number of PV and PQ buses, respectively. After solving for $x$, the remaining real power balance equation can be used to compute the generator real power injection at the slack bus. Similarly, the remaining $n_{p v}+1$ reactive power balance equations yield the generator reactive power injections.

\subsubsection{AC Power Flow Equations}\label{sec:ac_power_flow_equations}

These equations are derived from the real and reactive power relationships at each bus in the system, given the bus voltage magnitudes, phase angles, and the network parameters. In this derivation, we start with the complex power equations and develop the real and reactive power flow equations. In an AC power system, the complex power at bus $i$ is given by:
\begin{equation}
	S_i = P_i + jQ_i
\end{equation}
where:
\begin{itemize}
	\item $S_i$ is the complex power at bus $i$.
	\item $P_i$ is the real (active) power at bus $i$.
	\item $Q_i$ is the reactive power at bus $i$.
\end{itemize}
The complex power is related to the voltage at bus $i$ and the current injection at that bus as follows:
\begin{equation}
	S_i = V_i I_i^*
\end{equation}
where $V_i$ is the voltage at bus $i$, and $I_i^*$ is the complex conjugate of the current at bus $i$. Using Kirchhoff's Current Law (KCL) as described in Eq.~\ref{eq:kirchoff_current_low}, the current injection at bus $i$ can be written in terms of the bus admittance matrix, from Eq. :
\begin{equation}
	I_i = \sum_{j=1}^{n} Y_{ij} V_j
\end{equation}
where $Y_{ij}$ is the $(i,j)$ element of the bus admittance matrix, and $V_j$ is the voltage at bus $j$.
Substituting this expression into the equation for complex power, we get:
\begin{equation}
	S_i = V_i \left( \sum_{j=1}^{n} Y_{ij} V_j \right)^*
\end{equation}
Let the voltage at bus $i$ be written in polar form as:
\begin{equation}
	V_i = |V_i| e^{j\theta_i}
\end{equation}
where:
\begin{itemize}
	\item $|V_i|$ is the voltage magnitude at bus $i$.
	\item $\theta_i$ is the voltage phase angle at bus $i$.
\end{itemize}
Similarly, for bus $j$:
\begin{equation}
	V_j = |V_j| e^{j\theta_j}
\end{equation}
Now, substitute the polar forms of $V_i$ and $V_j$ into the complex power equation:
\begin{equation}
	S_i = |V_i| e^{j\theta_i} \left( \sum_{j=1}^{n} Y_{ij} |V_j| e^{j\theta_j} \right)^*
\end{equation}
Taking the complex conjugate:
\begin{equation}
	S_i = |V_i| e^{j\theta_i} \sum_{j=1}^{n} Y_{ij}^* |V_j| e^{-j\theta_j}
\end{equation}
This can be expanded as:
\begin{equation}
	S_i = |V_i| \sum_{j=1}^{n} |V_j| \left( G_{ij} \cos(\theta_i - \theta_j) + B_{ij} \sin(\theta_i - \theta_j) \right)
\end{equation}
where $G_{ij}$ and $B_{ij}$ are the conductance and susceptance between bus $i$ and bus $j$, respectively.

The real power $P_i$ at bus $i$ is the real part of the complex power $S_i$, while the reactive power $Q_i$ is the imaginary part. Therefore, the real power is given by:
\begin{equation}
	P_i = |V_i| \sum_{j=1}^{n} |V_j| \left( G_{ij} \cos(\theta_i - \theta_j) + B_{ij} \sin(\theta_i - \theta_j) \right)
\end{equation}
Similarly, the reactive power $Q_i$ is:
\begin{equation}
	Q_i = |V_i| \sum_{j=1}^{n} |V_j| \left( G_{ij} \sin(\theta_i - \theta_j) - B_{ij} \cos(\theta_i - \theta_j) \right)
\end{equation}
These are the real and reactive power flow equations at bus $i$ in an AC power system.

\subsection{Optimal Power Flow}
Optimal Power Flow (OPF) is a critical tool used in power system operations to ensure the efficient, stable, and reliable delivery of electricity. OPF aims to determine the optimal settings of control variables, such as generator outputs and tap changer positions, while ensuring the safe operation of the power system. Grid modeling is an essential component of OPF, as it provides the necessary mathematical framework for analyzing how electricity moves through the network. This article explores the primary objectives of grid modeling in the context of OPF, focusing on ensuring voltage and frequency stability, optimizing power flow, reducing losses, evaluating the integration of renewable energy, and forecasting load and generation changes.

\subsubsection{Ensure Voltage and Frequency Stability}

One of the primary objectives of grid modeling in OPF is to maintain voltage and frequency stability across the system. Voltage stability is necessary to ensure that all loads receive electricity within acceptable voltage limits, while frequency stability is crucial for synchronizing all generators to a common grid frequency, typically 50 Hz or 60 Hz.

In OPF, the system’s voltage stability is enforced by ensuring that the bus voltage magnitudes remain within specified limits:

\begin{equation}
	V_{\text{min}} \leq V_i \leq V_{\text{max}}, \quad i \in \mathcal{N}
\end{equation}

where $V_i$ is the voltage magnitude at bus $i$, and $\mathcal{N}$ is the set of all buses in the network. Frequency stability is ensured by balancing the total generation and load in the system, which is expressed as:

\begin{equation}
	\sum_{i \in \mathcal{G}} P_{g_i} = \sum_{i \in \mathcal{N}} P_{d_i} + P_{\text{loss}}
\end{equation}

where $P_{g_i}$ is the generation at bus $i$, $P_{d_i}$ is the demand at bus $i$, and $P_{\text{loss}}$ represents the total power losses in the network.

Grid modeling helps ensure that both voltage and frequency stability are maintained under various operating conditions by incorporating these constraints into the OPF problem.

\subsubsection{Optimise Power Flow and Reduce Losses}

Another key objective of grid modeling in OPF is to optimize power flow to minimize losses and operating costs. Power losses occur primarily due to the resistance in transmission lines, and these losses increase with the square of the current. The active power losses in a line connecting bus $i$ and bus $j$ can be approximated as:

\begin{equation}
	P_{\text{loss}, ij} = G_{ij} \left( V_i^2 + V_j^2 - 2 V_i V_j \cos(\theta_i - \theta_j) \right)
\end{equation}

where $G_{ij}$ is the conductance of the line, $V_i$ and $V_j$ are the voltage magnitudes at buses $i$ and $j$, and $\theta_i$ and $\theta_j$ are the voltage phase angles at those buses.

The objective function in OPF often seeks to minimize the total power losses or the operating cost of generation, which is typically modeled as:

\begin{equation}
	\min \sum_{i \in \mathcal{G}} C_i(P_{g_i})
\end{equation}

where $C_i(P_{g_i})$ is the cost function of the generator at bus $i$. Grid models in OPF incorporate line admittances, voltage constraints, and generation cost functions to determine the most efficient power flow pattern that minimizes losses and costs while satisfying all system constraints.

\subsection{Load Flow Algorithms for AC Power Flow}

There are various load flow algorithms that are developed to solve the AC power flow problem. These include:

\begin{itemize}
\item \textbf{Gauss-Seidel (GS) Method:} The Gauss-Seidel method is one of the simplest iterative techniques used to solve the power flow problem. It works by sequentially updating the voltage at each bus, assuming that the previously computed values for other buses are accurate. The method is easy to implement and has low computational requirements, making it suitable for small systems or as an initial solution for more complex algorithms. However, the convergence is generally slow, especially in large or heavily meshed networks. Despite its limitations, it can be useful in scenarios where computational resources are limited, or as a foundation for understanding more advanced methods.

\item \textbf{Newton-Raphson (NR) Method:} The Newton-Raphson method is a robust numerical approach that utilizes Taylor series expansion to solve the nonlinear algebraic equations of the power flow problem. It calculates the Jacobian matrix, which represents the sensitivity of power flow to voltage changes, and updates the bus voltages iteratively. This method offers faster convergence compared to Gauss-Seidel, especially for large-scale power systems. The Newton-Raphson method is widely used in modern power systems due to its high accuracy and ability to handle large networks. However, it is computationally intensive and requires more memory due to the formation and inversion of the Jacobian matrix.

\item \textbf{Fast Decoupled Load Flow (FDLF) Method:} The Fast Decoupled Load Flow method is an optimized version of the Newton-Raphson algorithm, designed to reduce computational complexity by exploiting the weak coupling between active and reactive power in most power systems. This leads to a simplified and decoupled Jacobian matrix, resulting in faster computation and reduced memory requirements. FDLF is highly efficient for real-time applications, such as online monitoring and control in energy management systems. However, its assumptions limit its accuracy in networks with strong coupling or in heavily loaded conditions, where the decoupling assumptions may not hold.

\item \textbf{DC Load Flow:} The DC load flow is a linearized approximation of the AC power flow problem. It assumes a lossless network, small voltage angle differences, and ignores reactive power and voltage magnitudes. This makes it computationally efficient, allowing for quick solutions, especially in scenarios such as contingency analysis, optimization studies, and transmission planning. While it provides useful approximations in high-voltage transmission networks, the DC load flow is not suitable for detailed studies of AC systems where voltage magnitude and reactive power play significant roles.

\item \textbf{Holomorphic Embedding Load Flow (HELM):} The Holomorphic Embedding Load Flow method is a relatively new approach based on complex analysis. It embeds the power flow problem into a holomorphic space, transforming it into a system that can be solved analytically, ensuring global convergence to the correct solution. HELM is particularly useful for solving power flow problems in ill-conditioned networks, where traditional methods like Newton-Raphson may fail to converge. Its guaranteed convergence makes it a promising tool for highly stressed systems, although its adoption in practical applications is still growing.

\item \textbf{Continuation Power Flow (CPF):} The Continuation Power Flow
is designed to analyze power systems under increasing load or stressed conditions. It traces the power flow solution as a loading factor is varied, enabling the identification of voltage stability limits and maximum loadability points. CPF is widely used in voltage stability studies and in determining the potential for voltage collapse in stressed power systems. Its ability to provide insights into system stability makes it a valuable tool for planning and reliability analysis, especially in systems nearing their operational limits.
	
\end{itemize}
	
\section{Types of Grid Models and Power Flow Analysis}

Power systems are complex, and ensuring their reliable and stable operation requires different types of grid models tailored to specific analyses. Power flow, or load flow, analysis is central to understanding how electricity moves through a network under steady-state conditions. However, understanding the full behavior of a power system requires not only steady-state models but also dynamic and frequency-domain models. Each of these models serves a specific purpose in power system analysis and stability, and each has distinct formulations that relate back to the power flow problem. In this article, we explore three primary types of grid models—steady-state, dynamic, and frequency domain—and their relation to the power flow problem.

\subsection{Steady-State Models}

Steady-state models are the foundation of power flow analysis and focus on ensuring that generation meets demand under various load conditions. These models represent the system when it is operating under normal, balanced conditions, and they are critical for planning and operational decision-making.

The power flow problem is typically formulated using the following set of nonlinear algebraic equations:

\begin{equation}
	P_i = |V_i| \sum_{j=1}^{n} |V_j| \left( G_{ij} \cos(\theta_i - \theta_j) + B_{ij} \sin(\theta_i - \theta_j) \right)
\end{equation}
\begin{equation}
	Q_i = |V_i| \sum_{j=1}^{n} |V_j| \left( G_{ij} \sin(\theta_i - \theta_j) - B_{ij} \cos(\theta_i - \theta_j) \right)
\end{equation}

where $P_i$ and $Q_i$ represent the real and reactive power injections at bus $i$, $|V_i|$ is the voltage magnitude, $\theta_i$ is the phase angle, and $G_{ij}$ and $B_{ij}$ are the conductance and susceptance between buses $i$ and $j$.

In steady-state models, these equations are used to calculate the bus voltages, phase angles, and power flows through the transmission lines under different load and generation scenarios. This type of modeling is critical for determining system reliability, voltage stability, and the ability to serve loads under normal conditions.

\subsection{Dynamic Models}

While steady-state models are useful for long-term planning and operation, they do not capture the transient behavior of the system during disturbances such as faults, switching events, or sudden changes in load. Dynamic models are necessary to understand how the system behaves during these events and to ensure system stability after a disturbance.

Dynamic models incorporate differential equations that describe how system variables such as rotor angles, frequencies, and voltages change over time. The swing equation, which governs the dynamics of generator rotor angles, is often used:

\begin{equation}
	M \frac{d^2 \delta}{dt^2} + D \frac{d\delta}{dt} = P_m - P_e
\end{equation}

where:
\begin{itemize}
	\item $M$ is the inertia constant.
	\item $D$ is the damping constant.
	\item $\delta$ is the rotor angle.
	\item $P_m$ is the mechanical power input.
	\item $P_e$ is the electrical power output.
\end{itemize}

Incorporating dynamic models into the power flow problem helps analyze how the system will react to transient events. The power flow equations are coupled with differential equations, leading to a dynamic power flow model that allows for the simulation of faults, voltage recovery, and generator responses.

\subsection{Frequency Domain Models}

Frequency domain models focus on analyzing the frequency stability of a power system, which becomes increasingly important with the integration of renewable energy sources such as wind and solar. These resources often introduce frequency variability due to their intermittent nature. In frequency domain analysis, the behavior of the system is evaluated in terms of how it responds to frequency deviations.

The frequency deviation $\Delta f$ from the nominal value is governed by the balance between generation and load, represented by:

\begin{equation}
	\Delta f = \frac{1}{2 H} \left( P_{gen} - P_{load} - D \Delta f \right)
\end{equation}

where:
\begin{itemize}
	\item $H$ is the system's inertia constant.
	\item $P_{gen}$ is the generated power.
	\item $P_{load}$ is the load demand.
	\item $D$ is the frequency damping coefficient.
\end{itemize}

In frequency domain models, the power flow problem is extended to analyze how frequency deviations propagate through the system. The bus admittance matrix can be augmented to include frequency-dependent terms, and control mechanisms like primary frequency response and automatic generation control (AGC) are modeled to stabilize frequency under different conditions.

\subsection{Discussion}

Each type of grid model—steady-state, dynamic, and frequency domain extends the fundamental power flow problem to address different aspects of system behavior. The steady-state power flow equations provide the base for determining voltages, power flows, and losses under normal conditions. Dynamic models augment the power flow problem by adding time-dependent equations to analyze system response to disturbances, while frequency domain models modify the power flow problem to focus on frequency stability, particularly in the presence of renewable generation. In all cases, the power flow equations remain the core mathematical framework, but they are modified or extended with additional terms and constraints to analyze specific phenomena in the power system.

\section{Renewable Energy Sources Modeling}

Renewable energy sources, such as solar photovoltaic (PV), wind, hydroelectric, and geothermal, are increasingly integrated into modern power systems due to their sustainable nature and the global push for decarbonization. 
Unlike conventional fossil fuel-based energy sources, renewable energy sources naturally replenish over time and are environmentally friendly. The hydroelectric and geothermal energy sources are controllable due to persistent availability of their driving energy source. In comparison, PV and wind have intermittent nature and dependency on weather conditions present challenges in power flow modeling and grid stability analysis.

\subsection{Renewable Energy Sources}

\subsubsection{Solar Photovoltaic (PV)}
Solar PV systems convert sunlight into electricity using semiconductor materials, typically silicon. The output of a PV system depends on the intensity of solar radiation, which varies throughout the day and is influenced by weather conditions. Solar PV systems are often integrated into distribution networks, either as rooftop installations or larger solar farms.

\subsubsection{Wind Energy}
Wind turbines convert the kinetic energy of wind into mechanical energy, which is then transformed into electrical energy by a generator. The power output of a wind turbine is a function of wind speed, air density, and the turbine's characteristics. Wind energy is intermittent, and its integration into the grid requires advanced control strategies to handle fluctuations.

\subsubsection{Hydroelectric Energy}
Hydroelectric energy is generated by the flow of water through turbines, typically in dams or rivers. The availability of hydroelectric power depends on water flow, which is influenced by seasonal variations and rainfall. Hydroelectric power is a reliable and well-established renewable source that can provide both base-load and peaking power.

\subsubsection{Geothermal Energy}
Geothermal energy harnesses heat from beneath the Earth's surface to generate electricity. Steam from hot water reservoirs is used to drive turbines and generate power. Geothermal energy is location-dependent but offers a consistent and reliable source of renewable energy.

\subsection{Power Flow Problem with Renewable Energy Sources}

In power system analysis, the power flow (also known as load flow) problem determines the voltage magnitude and phase angle at each bus in a power network under steady-state conditions as discussed in Section \ref{section:power_flow}. The presence of renewable energy sources adds complexity to the traditional power flow problem due to their intermittent nature and variability.

The power flow problem is typically solved using the following set of nonlinear algebraic equations as derived in the Section \ref{sec:ac_power_flow_equations}:
\begin{equation}
	P_i = |V_i| \sum_{j=1}^{n} |V_j| \left( G_{ij} \cos(\theta_i - \theta_j) + B_{ij} \sin(\theta_i - \theta_j) \right)
\end{equation}
\begin{equation}
	Q_i = |V_i| \sum_{j=1}^{n} |V_j| \left( G_{ij} \sin(\theta_i - \theta_j) - B_{ij} \cos(\theta_i - \theta_j) \right)
\end{equation}

For renewable energy sources, the power injection terms $P_i$ and $Q_i$ will vary based on the availability of the renewable resource. For instance, solar PV output depends on the irradiance, while wind power depends on the wind speed. These stochastic behaviors can be modeled using probabilistic or time-series approaches to better capture the intermittent characteristics.

\subsubsection{Renewable Energy Source Modeling}
In the power flow problem, renewable energy sources are typically modeled as generator buses with variable output, depending on the availability of the resource. The active power output of a renewable generator at bus $i$, $P_{\text{RES}, i}$, can be expressed as:

\begin{equation}
	P_{\text{RES}, i} = P_{\text{max}, i} \cdot \rho_i
\end{equation}

where $P_{\text{max}, i}$ is the maximum power output of the renewable generator at bus $i$, and $\rho_i$ is a stochastic variable representing the availability of the resource (e.g., wind speed or solar irradiance).

Grid modeling in OPF helps evaluate how the integration of RES affects the overall power flow, ensuring that the system can accommodate the variability while maintaining stability and minimizing costs.

\subsubsection{Forecast Load and Generation Changes}

In modern power systems, accurate forecasting of load and generation is critical for efficient and reliable operations. Grid models in OPF are used to forecast future scenarios and adjust the system's control variables accordingly. Forecasting models predict changes in load demand and generation output, especially for renewable energy sources.

Load forecasting is typically represented in grid models as time-varying demand functions:

\begin{equation}
	P_{d_i}(t) = P_{d_i} \cdot f(t)
\end{equation}

where $P_{d_i}$ is the base demand at bus $i$, and $f(t)$ is a time-dependent function capturing the load variation over time.

Similarly, renewable generation forecasting takes into account expected variations in wind speed or solar irradiance over time, allowing for better scheduling of reserves and generation dispatch. Grid modeling in OPF incorporates these forecasts to create robust plans that can handle future uncertainties in both load and generation.

The integration of renewable energy sources (RES) such as wind and solar into power systems introduces significant uncertainty due to their inherent variability. Effectively modeling this uncertainty is critical for ensuring reliable and efficient grid operations. Two common approaches to address this challenge involve single-period and multi-period simulations, with a focus on deterministic and stochastic methods for handling forecast variables.

\subsection{Single-Period Simulation}
In a single-period simulation, renewable generation is modeled over a fixed, isolated time frame. This method is often used for short-term planning or operational decisions, such as determining energy dispatch or reserve requirements for a specific hour or day.

\subsubsection{Deterministic Handling}
In deterministic models, renewable generation forecasts are treated as known, fixed values. These forecasts are based on historical data, weather predictions, or engineering estimates. Although easy to implement, deterministic methods do not account for forecast errors, limiting their accuracy in real-world scenarios where weather and other factors can cause significant deviations.

\subsubsection{Stochastic Handling}
Stochastic models, on the other hand, explicitly account for the uncertainty in renewable forecasts by treating generation as a random variable with a probability distribution. Techniques such as Monte Carlo simulations or scenario generation are used to model a range of possible outcomes. This allows for more robust decision-making by considering the probability of under- or over-generation.

\subsection{Multi-Period Simulation}
Multi-period simulation extends the analysis over multiple time steps, allowing for the modeling of renewable generation's temporal evolution. This is especially important for long-term operational decisions like energy storage optimization or reserve scheduling, where inter-temporal relationships are key.

\subsubsection{Deterministic Handling} 
Similar to single-period models, multi-period deterministic simulations assume known forecasts for each time step. While this approach provides a basic understanding of generation patterns, it risks mismanagement of resources if actual conditions deviate from the forecast, especially over longer time horizons.

\subsubsection{Stochastic Handling}
Stochastic multi-period simulations introduce randomness across time periods, accounting for both intra-day and day-ahead uncertainties. These models are often formulated as stochastic optimization problems, using methods like stochastic dynamic programming to capture the sequential decision-making process under uncertainty. They enable planners to assess the impact of varying renewable output over time and optimize decisions accordingly.

\subsection{Discussion}
Modeling uncertainty in renewable generation requires careful consideration of both the time horizon (single-period vs. multi-period) and the handling of forecast variables (deterministic vs. stochastic). While deterministic models offer simplicity, stochastic methods provide a more realistic representation of uncertainties, helping to enhance grid reliability in the face of fluctuating renewable energy outputs.

\section{Use-Case Based Analysis}
\begin{figure} [htb]
	\centering
	\includegraphics[trim=0cm 0cm 0cm 0cm, width=0.6\textwidth]{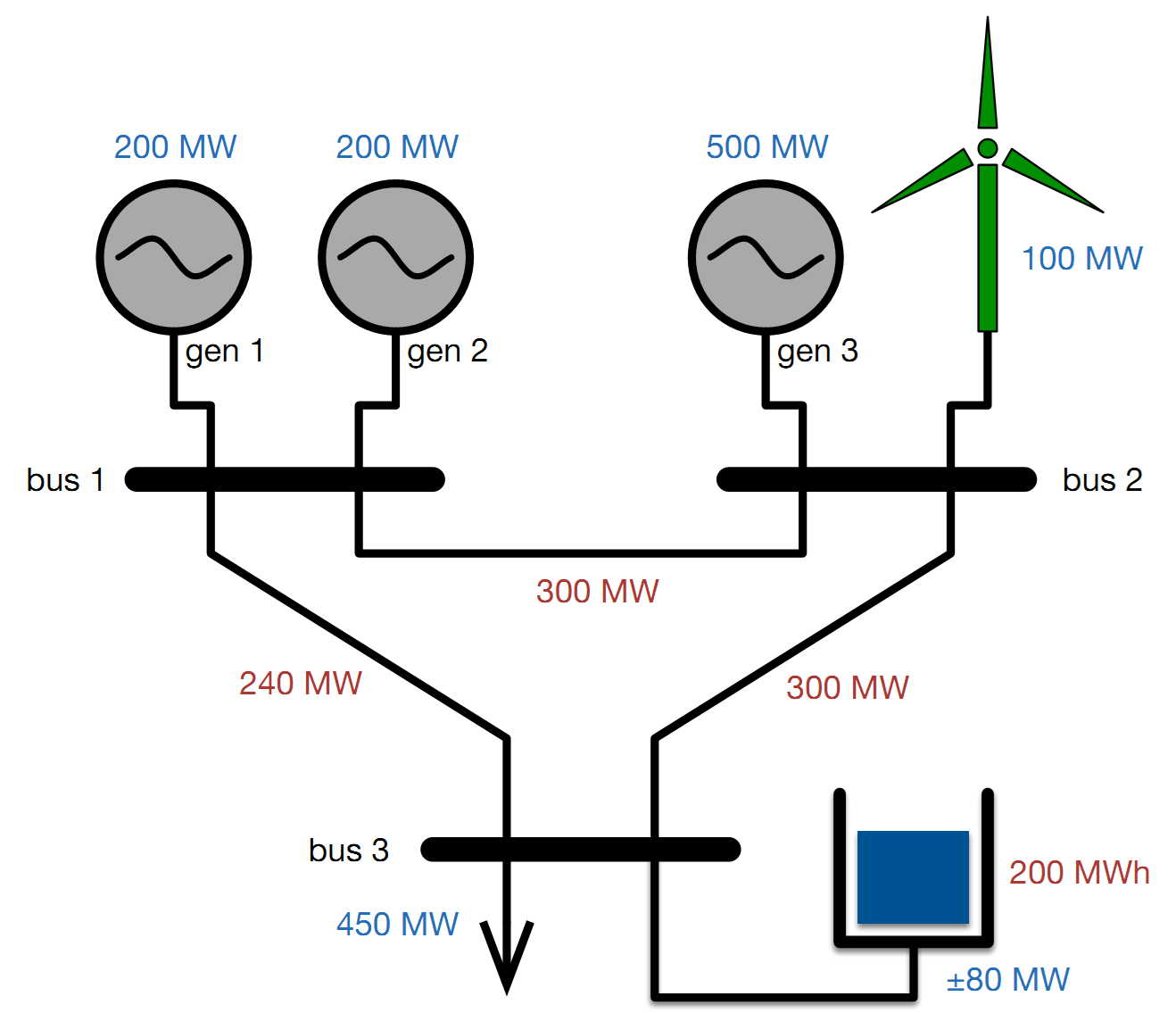}
	\caption{Use-case example.}
	\label{figure: use_case}
\end{figure}

Consider the network shown in Fig.\ref{figure: use_case}. This network can be simulated in Matpower using Matpower Optimal Scheduling Tool (MOST) \cite{lamadrid2018using}.  

Following is the brief description of the network:
\begin{itemize}
	\item Two identical 200 MW gens at bus 1 with different reserve cost and a 500 MW generator at bus 2.
	\item 450 MW at bus 3 curtailable at \$1000/MWh
	\item 300 MW limit of line 1–2, 240 MW limit of line 1–3 and 300 MW limit of line 2–3
	\item Wind unit at bus 2 with 100 MW output in nominal case
	\item Storage of 200 MWh unit at bus 3 with 80 MW max charge/discharge rate
\end{itemize}

The AC optimal power flow and DC optimal power flow for a single period with all deterministic values gives a base line economic dispatch results.
\subsection{Simulation of Single Point in Time}
Following is the code in Matpower that simulates the AC optimal power flow for single point in time.
\begin{lstlisting}[caption=AC optimal flow for a single period deterministic case]
mpc = loadcase("ex_case3a");
r1 = runopf(mpc, mpopt);
\end{lstlisting}
The result is shown below:
\begin{table}[htb]
	\centering
	\caption{Bus data of the use-case.}
	\begin{tabular}{|l|l|l|l|l}
		\cline{1-4}
		Bus ID & Bus Type & PD  & QD &  \\ \cline{1-4}
		1      & 3        & 0   & 0  &  \\ \cline{1-4}
		2      & 2        & 0   & 0  &  \\ \cline{1-4}
		3      & 2        & 450 & 0  &  \\ \cline{1-4}
	\end{tabular}
\end{table}

\begin{table}[htb]
	\centering
	\caption{Generator dispatch data of the use-case.}
	\begin{tabular}{|l|l|l|}
		\hline
		Gen\_Bus & PG   & QG \\ \hline
		1        & 575  & 0  \\ \hline
		1        & 125  & 0  \\ \hline
		2        & 200  & 0  \\ \hline
		3        & -450 & 0  \\ \hline
	\end{tabular}
\end{table}
More information can be obtained from MATPOWER and MOST documentation.

\subsection{Multi-period simulation with deterministic forecast}

\begin{figure} [h]
	\centering
	\includegraphics[trim=0cm 0cm 0cm 0cm, width=0.7\textwidth]{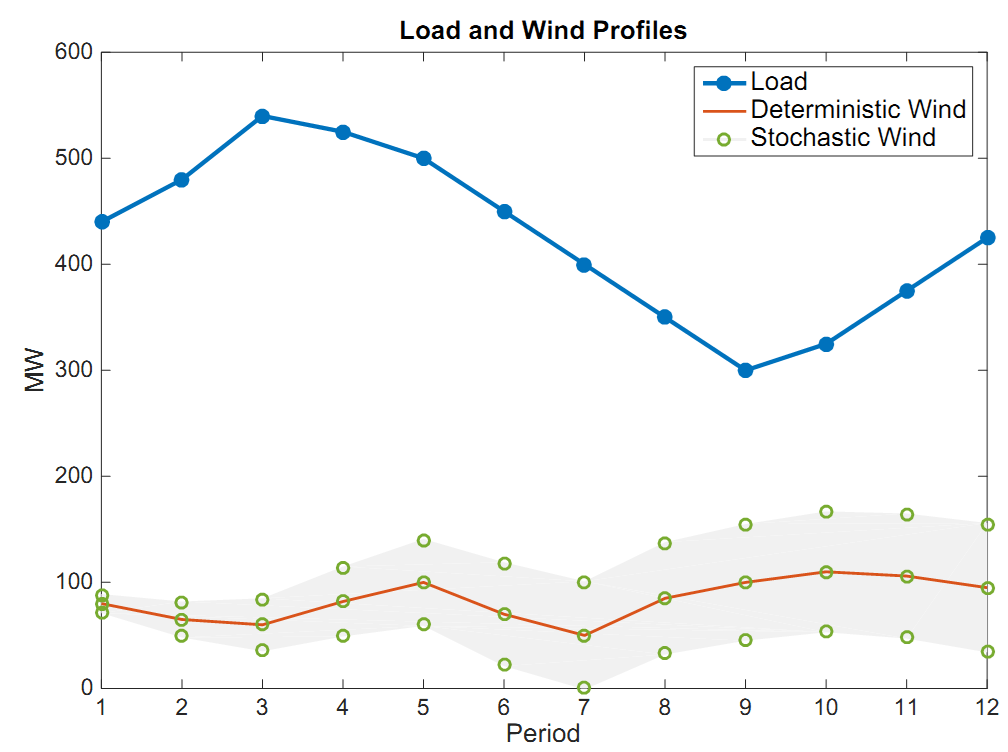}
	\caption{Multi-period deterministic forecast of energy demand and wind power.}
	\label{figure: use_case_multiperiod_deterministic}
\end{figure}

\begin{figure} [htb]
	\centering
	\includegraphics[trim=0cm 0cm 0cm 0cm, width=0.8\textwidth]{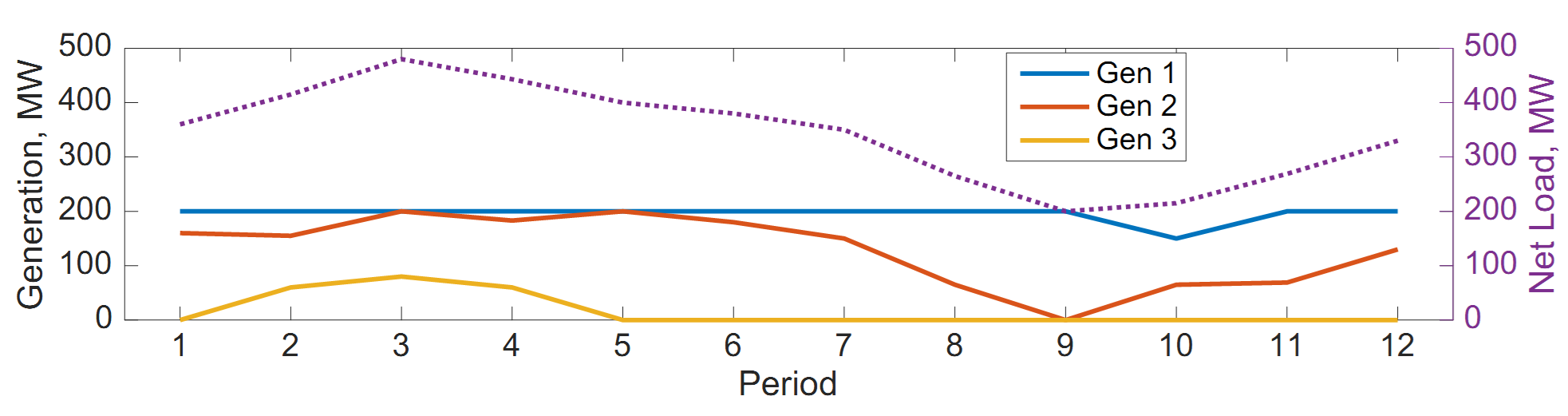}
	\caption{Generation dispatch for multi-period simulation considering the deterministic wind forecast.}
	\label{figure: use_case_multiperiod_deterministic_dispatch}
\end{figure}

The total load active power connected to bus 3 varies for the duration of 12 hours as shown in the figure \ref{figure: use_case_multiperiod_deterministic}. The wind forecast shows that uncertainty in the forecast shown as red-line increases over time. However for the simulation sake, only the red-line based wind power is considered. The result of the dispatch of generation for the baseline wind forecast shown in red color is shown in the figure~\ref{figure: use_case_multiperiod_deterministic_dispatch}.

\subsection{Multi-period simulation with stochastic forecast as scenarios}
\begin{figure} [htb]
	\centering
	\includegraphics[trim=0cm 0cm 0cm 0cm, width=0.6\textwidth]{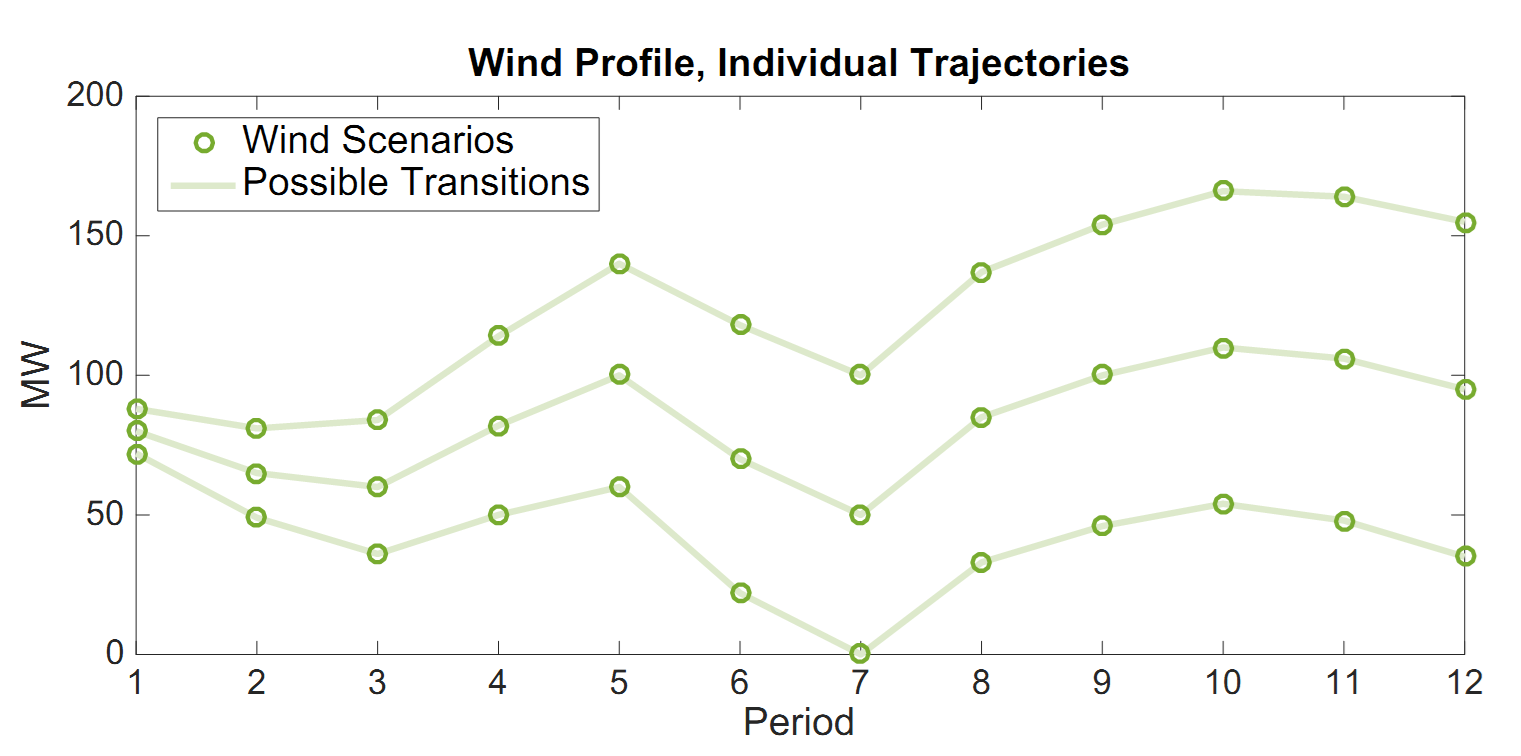}
	\caption{Stochastic modeling of wind power production as scenarios.}
	\label{figure: use_case_multiperiod_stochastic}
\end{figure}

\begin{figure} [htb]
	\centering
	\includegraphics[trim=0cm 0cm 0cm 0cm, width=0.8\textwidth]{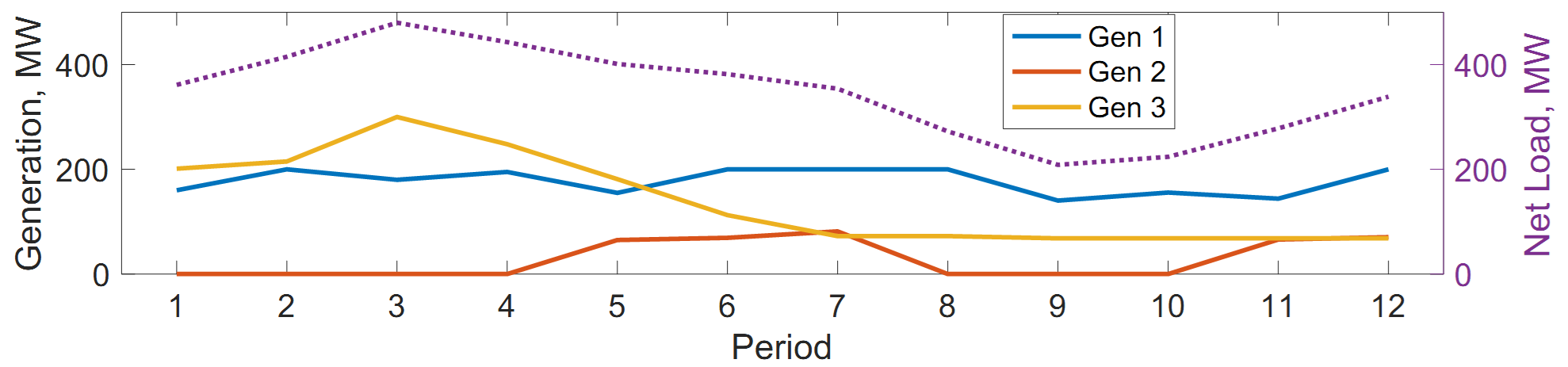}
	\caption{Stochastic dispatch scenario for the multi-period wind forecast considering worst case analysis.}
	\label{figure: stochastic_dispatch_scenario}
\end{figure}

The stochastic forecast can be considered as scenarios as shown in figure \ref{figure: use_case_multiperiod_stochastic}.
The scenarios are the outcome of sampling of transition probability matrices with an assumption that if the system is in the high wind state in the first period, it
will stay in the high wind state in every subsequent period, and the same with the
average and low wind states. 
These scenarios will lead to dispatch possibilities for each case. One of the dispatch result is shown in figure \ref{figure: stochastic_dispatch_scenario}.

\subsection{Multi-period simulation with stochastic forecast as full transition probabilities}
\begin{figure} [htb]
	\centering
	\includegraphics[trim=0cm 0cm 0cm 0cm, width=0.6\textwidth]{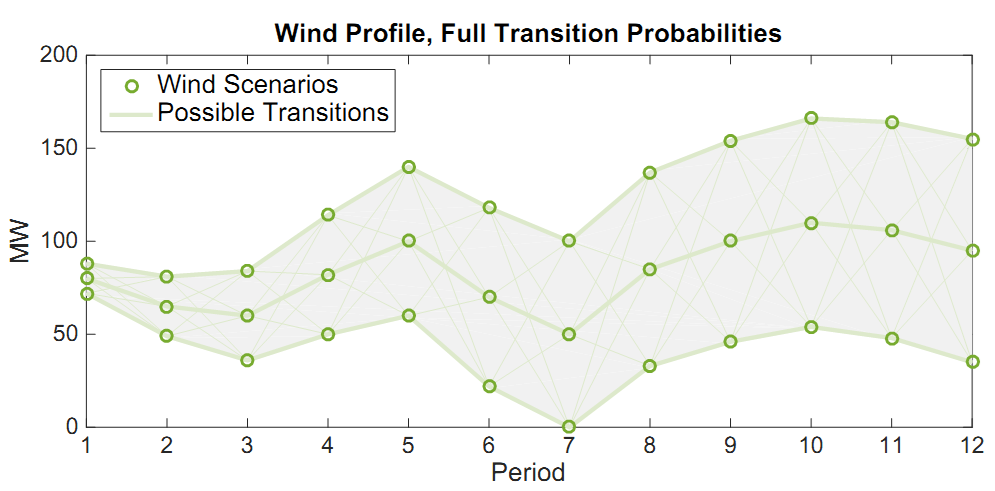}
	\caption{Wind forecast modeled as transition probability matrix.}
	\label{figure: use_case_multiperiod_stochastic_full_transition}
\end{figure}

\begin{figure} [htb]
	\centering
	\includegraphics[trim=0cm 0cm 0cm 0cm, width=0.8\textwidth]{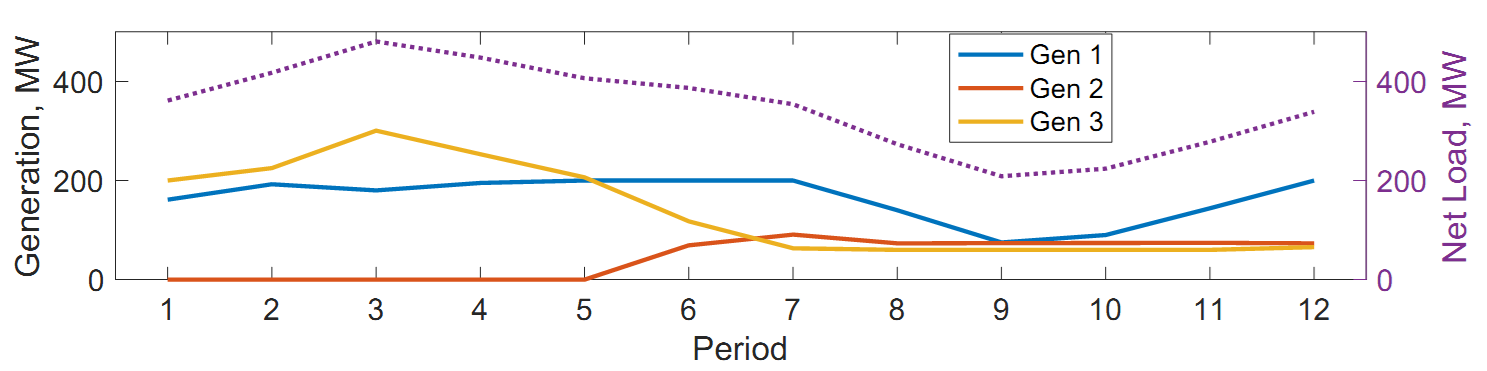}
	\caption{Stochastic dispatch scenario for the multi-period wind forecast considering transition probabilities.}
	\label{figure: dispatch_schedule_multi_period_stochastic_transition_probab_mat}
\end{figure}

The more general case of stochastic dispatch uses the full transition probability matrices, where transitions between low, average and high wind scenarios are allowed from period to period as illustrated in figure \ref{figure: use_case_multiperiod_stochastic_full_transition}. Both the cases shall simulate three scenarios, the peak-max, peak-min and the average trend. However, the only difference is that the higher ram rate requirements will be considered while transiting from the lower values to higher values. This in turn shall impact the dispatch schedule as shown in figure \ref{figure: dispatch_schedule_multi_period_stochastic_transition_probab_mat}.

\subsection{Discussion}
From the analysis we can observe that uncertainty in the wind power production can significantly impact the power dispatch schedules of the generation. 
This situation need to handle the ramp-rate constraints, reserve requirements and dispatch constraints as well.

\newpage
\section{Further Reading}

There are also further two key aspects related to the grid modelling of renewable energy: \textit{Power Quality} and \textit{Grid Stability}, and examines solutions that address the challenges posed by renewable energy sources.
	
\subsection{Power Quality}

Renewable energy systems, especially intermittent sources like wind and solar, can introduce several power quality issues, including voltage fluctuations, harmonic distortion, and frequency deviations. The variable nature of renewable energy sources can cause rapid changes in the voltage levels, impacting sensitive equipment and creating grid instability. Additionally, the use of inverters in renewable systems introduces harmonic currents, which distort the sinusoidal waveform of voltage and current, leading to operational issues in electronic equipment and increased system losses. Furthermore, the asynchronous generation of renewables can result in frequency deviations from the nominal grid frequency, adversely affecting frequency-sensitive devices. To mitigate these power quality issues, advanced inverter designs and control strategies are employed, such as harmonic filters and active power control systems, which maintain voltage and current waveform integrity. For further reading, see \cite{bollen2011integration}.

\subsection{Grid Stability}

Grid stability is traditionally supported by the inertia provided by the rotating mass of conventional power plants. However, renewable energy sources like wind and solar, being inverter-based, do not contribute the same level of inertia, creating challenges for grid frequency stability. To address this, modern inverters can provide synthetic inertia by adjusting output power in response to frequency changes, mitigating frequency deviations and improving grid stability. Grid-forming inverters, which operate as virtual synchronous machines, mimic the behavior of traditional generators by adjusting their output based on real-time grid conditions, thereby stabilizing frequency and ensuring a reliable power supply. Additionally, inverters can be programmed to control reactive power, which is crucial for maintaining stable voltage levels across the grid, especially during fluctuations in renewable generation. Battery Energy Storage Systems (BESS) also play a vital role by providing fast frequency response, stabilizing both voltage and frequency, and storing excess renewable energy to be discharged during periods of low generation or high demand. For further reading, see \cite{vittal2019power}.

\newpage
\bibliography{library}

\end{document}